\documentstyle[12pt,epsfig]{article}

\newcommand{\be}{\begin{equation}}
\newcommand{\ee}{\end{equation}}
\def\aprle{\buildrel < \over {_{\sim}}}
\def\aprge{\buildrel > \over {_{\sim}}}
\begin{document}
\topmargin 0pt
\oddsidemargin=-0.4truecm
\evensidemargin=-0.4truecm
\renewcommand{\thefootnote}{\fnsymbol{footnote}}
\newpage
\setcounter{page}{1}
\begin{titlepage}     
\vspace*{-2.0cm}  
\begin{flushright}
FISIST/16-2000/CFIF \\
hep-ph/0011136
\end{flushright}
\vspace*{0.5cm}
\begin{center}
\vspace*{0.2cm}
{\Large \bf Matter effects in oscillations of neutrinos traveling \\
\vspace{0.1cm}
short distances in matter}
\\
\vspace{1.0cm}

{\large E. Kh. Akhmedov
\footnote{On leave from National Research Centre Kurchatov Institute, 
Moscow 123182, Russia. E-mail: akhmedov@cfif.ist.utl.pt}}\\
\vspace{0.05cm}
{\em Centro de F\'\i sica das Interac\c c\~oes Fundamentais (CFIF)} \\
{\em Departamento de F\'\i sica, Instituto Superior T\'ecnico}\\
{\em Av. Rovisco Pais, P-1049-001 Lisboa, Portugal}\\
\end{center}
\vglue 1.2truecm

\begin{abstract} 
It is well known that when the distance $t$ traveled by neutrinos in 
matter is short, matter effects in oscillations of neutrino flavour are 
small and decrease with decreasing $t$ more rapidly than the 
oscillations effects themselves. We discuss the reason for this and 
demonstrate that under certain circumstances this statement is no longer
correct. In particular, we show that if neutrinos propagate significant 
distances in vacuum before entering matter (or after exiting it), matter 
effects in short-$t$ neutrino oscillations can be significantly enhanced. 
Implications for oscillations of solar and atmospheric neutrinos with 
nearly horizontal trajectories inside the earth and for 
accelerator experiments are considered. We also comment on neutrino 
oscillations in matter due to flavour changing neutral currents. 
\end{abstract}
\vspace{1.cm}
\centerline{Pacs numbers: 14.60.+Pq, 26.65.+t} 
\vspace{.3cm}
\centerline{Keywords: neutrino oscillations, matter effects} 
\end{titlepage}
\renewcommand{\thefootnote}{\arabic{footnote}}
\setcounter{footnote}{0}
\newpage
\section{Introduction}
It is well known that when the distance $t$ traveled by neutrinos in 
matter is short, the probabilities of oscillations of neutrino flavour in
matter reduce to those in vacuum, i.e. with decreasing $t$ matter 
effects on neutrino oscillations die out more rapidly than the oscillations 
effects themselves. 
In the present Letter we show that under certain conditions this is no 
longer true and matter effects in short-$t$ experiments can be 
quite significant or even dominate the oscillation probability. 
In particular, we show that if neutrinos propagate significant distances in 
vacuum before entering matter (or after exiting it) matter effects 
in short-$t$ neutrino oscillations can be strongly enhanced.
We also discuss oscillations of neutrino mass eigenstates in matter, 
which are relevant for oscillations of solar and supernova neutrinos 
inside the earth. We show that these oscillations can have sizeable
probabilities even when neutrino pathlengths in matter are relatively short. 

\section{Two-flavour neutrino evolution in matter}

Consider the evolution equation for two-flavour neutrino oscillations 
in matter in the weak eigenstate basis \cite{W,MS}: 
\be
i\frac{\partial}{\partial t}
\left(\begin{array}{c}
\nu_a \\
\nu_b
\end{array}\right)=
\left(\begin{array}{cc}
-A(t) & B \\
B & A(t) \end{array} \right)\left(\begin{array}{c}
\nu_a \\
\nu_b
\end{array}\right)\equiv H(t)\left(\begin{array}{c}
\nu_a \\
\nu_b   
\end{array}\right)
\label{Sch}
\ee
Here $\nu_{a,b}(t)$ are the probability amplitudes of finding neutrinos 
of the corresponding flavor $a, b$ at a time $t$ (in particular, one of
these two species can be a sterile neutrino $\nu_s$). The parameters $A$ 
and $B$  are  
\be
A(t)=\cos 2\theta_0\,\delta-V(t)\,,\quad\quad
B=
\sin 2\theta_0\,\delta\,,
\label{AB}
\ee
where
\be
\delta = \frac{\Delta m^2}{4E}\,,\quad\quad 
V(t)=\frac{G_F}{\sqrt{2}}\,N(t)\,. 
\label{Vdelta}
\ee
Here $G_F$ is the Fermi constant, $E$ is neutrino energy, $\Delta
m^2=m_2^2-m_1^2$, where $m_{1,2}$ are the neutrino mass eigenvalues, and
$\theta_0$ is the mixing angle in vacuum. The effective density $N(t)$
depends on the type of the neutrinos taking part in the oscillations. 
For transitions between antineutrinos one should substitute $-N$ for $N$
in eq. (\ref{AB}). 
In a matter of constant density the probability of $\nu_a\leftrightarrow 
\nu_b$ oscillations takes a very simple form
\be
P(\nu_a\to \nu_b;\, t)=\sin^2 2\theta \sin^2 \omega t\,,
\label{P1}
\ee
where $\theta$ and $2\omega$ are the mixing angle in matter and the energy  
splitting between the matter eigenstates respectively:
\be
\sin 2\theta=\frac{B}{\omega}=\sin 2\theta_0 \frac{\delta}{\omega}\,, 
\quad\quad \omega=\sqrt{A^2+B^2}\,.
\label{param1}
\ee
It is now easy to see that in the short baseline limit ($\omega t\ll 1$)
the oscillation probability 
\be
P(\nu_a\to \nu_b;\,t)\simeq \sin^2 2\theta\,(\omega t)^2 =\sin^2 2\theta_0
\,(\delta \!\cdot\! t)^2
=\sin^2 2\theta_0 \,(\frac{\Delta m^2}{4E} t)^2 \,,
\label{P2}
\ee
which is just the short-baseline limit of the oscillation probability 
in vacuum \cite{vaclimit}.

This raises a number of questions:

$\bullet$ Why do the matter effects on oscillations die out with decreasing 
$t$ more rapidly than the oscillation effects themselves? 

$\bullet$ Is this true also in a matter of non-constant density?

$\bullet$  Is this true in the case of oscillations between more than two
neutrino species?

$\bullet$  Are there any conceivable situations when the matter effects 
on oscillations of neutrinos traveling short distances in matter can be as
large as the oscillation effects themselves?

The answer to the first of these questions will help us to find the 
answers to the rest of them.

\section{Short-pathlength neutrino oscillations in matter}

Consider the two-flavour evolution equation (\ref{Sch}) in the short
baseline (small oscillation phase) limit. In this limit the oscillation 
effects are known to be small, so one can solve the evolution equation in
perturbation theory. Assume that the initial neutrino state is $\nu_a$. 
Then in the leading order in perturbation theory the amplitude 
$\nu_b(t)=-iBt$, which immediately yields (\ref{P2}). Thus, the fact that 
the matter effects on neutrino flavour oscillations disappear in the short 
baseline limit is the consequence of 
the two facts: (1) the oscillations effects 
are small and the perturbation theory applies; (2) since the initial 
state is a flavour eigenstate, in the leading order the transition amplitude 
is only determined by the off-diagonal terms in the effective Hamiltonian in 
(\ref{Sch}), whereas the matter effects enter only through the diagonal terms. 

{}From the above the answers to the other questions that we asked immediately 
follow: 
The functional $t$ dependence of matter density is irrelevant, i.e. our
conclusion holds irrespective of whether the matter density is constant or 
not; it is also true in the case of oscillations between more than two
species because, in the weak eigenstate basis, matter 
density enters only into the diagonal terms of the effective Hamiltonian 
for any number of flavours; the above conclusions need not be correct if the 
initial neutrino state is not a weak eigenstate, i.e. not a neutrino of a
definite flavour. 

Let us now consider the situation when the initial neutrino state entering
the matter is a coherent superposition of the flavour eigenstates $\nu_a$ 
and $\nu_b$:
$|\nu_i\rangle = |\nu(0)\rangle=a_1|\nu_a\rangle+a_2|\nu_b\rangle$, 
$|a_1|^2+|a_2|^2=1$. 
In what follows, unless otherwise is specified, we shall always understand
by neutrino pathlength the distance traveled by neutrinos {\it in matter}. 
It is straightforward to find the transition
probabilities to the second order in perturbation theory: 
\be
P(\nu_i\to \nu_a; \,t)\simeq |a_1|^2-(|a_1|^2-|a_2|^2)(B t)^2 -
2 \mbox{Im}(a_2^* a_1) (B t)-4 \mbox{Re}(a_2^*a_1) B\int_0^t dt_1
\int_0^{t_1} dt_2 A(t_2)\,,
\label{P3}
\ee
\be
P(\nu_i\to \nu_b;\, t)\simeq |a_2|^2+(|a_1|^2-|a_2|^2)(B t)^2 + 
2 \mbox{Im}(a_2^* a_1) (B t)+4 \mbox{Re}(a_2^*a_1) B\int_0^t dt_1
\int_0^{t_1} dt_2 A(t_2)\,,
\label{P4}
\ee
For our further discussion it is useful to write down the simplified
versions of these expressions relevant for neutrino oscillations in a 
matter of constant density: 
\be
P(\nu_i\to \nu_a; \,t)\simeq |a_1|^2-(|a_1|^2-|a_2|^2)(B t)^2 -
2 \mbox{Im}(a_2^* a_1) (B t)-2 \mbox{Re}(a_2^*a_1) (A B \,t^2)\,,
\label{P3a}
\ee
\be
~~P(\nu_i\to \nu_b; \,t)\simeq |a_2|^2+(|a_1|^2-|a_2|^2)(B t)^2 + 
2 \mbox{Im}(a_2^* a_1) (B t)+2 \mbox{Re}(a_2^*a_1) (A B \,t^2)\,.
\label{P4a}
\ee
Matter effects enter through the parameters $A$ in the last terms on the 
r.h.s. of eqs. (\ref{P3})-(\ref{P4a}). 
Notice that the leading order matter-induced contribution is $\sim V\!\delta\! 
\cdot\!t^2$, whereas in the case when the initial state is a pure flavour 
eigenstate the leading order matter effect is $\sim (V\!\delta\!\cdot\!t^2)^2$ 
(assuming $V\aprge\delta$). 
Thus, matter effects in short-pathlength neutrino oscillations can be strongly 
enhanced when the initial state is not a pure flavour eigenstate
\footnote{It is also important that the term $V\!\delta\!\cdot\!t^2$ is 
sensitive to the sign of $\Delta m^2$, whereas $(V\!\delta\!\cdot\!t^2)^2$
is not.}. 

As can be seen from (\ref{P3})-(\ref{P4a}), the oscillation effects themselves 
get enhanced in this case provided that $\mbox{Im}(a_2^* a_1)\ne 0$, the 
leading contribution to the transition probability being now $\sim\delta\! 
\cdot\!t$ rather than $\sim (\delta\!\cdot\!t)^2$. 
 
\section{Two-media neutrino oscillations}

How can one create an initial neutrino state which is not a pure flavour 
eigenstate? One possibility is to let neutrinos oscillate in a different 
medium before entering the medium of interest. For example, if neutrinos 
propagate in vacuum before they enter the matter, the initial state 
arriving at the vacuum-matter border is no longer a flavour eigenstate but 
rather a coherent superposition of the flavour eigenstates. 

Assume that neutrinos are initially produced in the flavour eigenstate
$\nu_a$, propagate a distance $t_1$ in vacuum and then the resulting state 
propagates a distance $t$ in a matter of constant density. The oscillation 
phase acquired in vacuum is $\phi_1=\delta\!\cdot\! t_1$. Let us denote 
$\sin\phi_1=s_1$, $\cos\phi_1=c_1$. Then the state $|\nu_i\rangle$ entering 
the matter is characterized by $a_1=c_1+i \cos 2\theta_0\, s_1$, $a_2=-i 
\sin 2\theta_0\, s_1$. In the limit of short-pathlength oscillations in
matter one finds from (\ref{P3a}) and (\ref{P4a})
\footnote{These probabilities can also be directly obtained (as a small 
$\phi_2$ limit) from the general expression for the evolution matrix 
for neutrino oscillations in two layers of different constant densities 
derived in \cite{Akh}.}
\be
P(\nu_i\to \nu_b; \,t)\simeq \sin^2 2\theta_0\left\{s_1^2+(c_1^2 - s_1^2) 
(\delta\!\cdot\!t)^2 + 2 s_1 c_1 (\delta\!\cdot\!t) + 2 s_1^2 \cos 
2\theta_0(V\!\delta\!\cdot\!t^2)\right\} 
\label{P5}
\ee
and $P(\nu_i\to \nu_a; \,t)=1-P(\nu_i\to \nu_b;\, t)$. Notice that the
l.h.s. of eq. (\ref{P5}) can also be understood as $P(\nu_a\to \nu_b;
\,t_1+t)$.  

Eq. (\ref{P5}) has a simple physical interpretation. In the limit $t\to 0$ 
one has $P(\nu_i\to \nu_b; \,t)=\sin^2 2\theta_0\,s_1^2$ which corresponds 
to propagation only in vacuum; the second and the third terms in the curly 
brackets are due to the increase of the oscillation phase during the time
interval $t$, neglecting the matter effects. These terms come from the 
expansion of $\sin^2 (\phi_1+\delta\!\cdot\!t)$ in small $\delta\!\cdot\!t$. 
The fourth term is the leading order matter contribution. Notice that 
we do not assume the smallness of the phase $\phi_1$, and in fact a sizeable 
enhancement of matter effects is only possible when it is not small. 

Several comments are in order.

(i) The leading-order matter contribution vanishes in the case of maximal 
mixing in vacuum, $\theta_0=\pi/4$. This may be useful for studying
deviations of lepton mixing from the maximal one. 

(ii) The first three terms in the curly brackets in eq. (\ref{P5}) are even 
in $\delta$ while the last, matter-induced, term is odd in it and so is
sensitive to the sign of $\Delta m^2$. 
\footnote{More precisely, it is sensitive to the sign of $\cos 2\theta_0\,
\Delta m^2$ as only the sign of this quantity has a physical meaning. 
We adopt the convention $\cos2\theta_0\ge 0$ and allow for both positive
and negative signs of $\Delta m^2$.}
Thus, enhanced matter effects can 
facilitate studying the type of the neutrino mass hierarchy. 

(iii) Although the absolute magnitude of matter effects, of course, strongly 
depends on $\theta_0$, their relative size is only mildly $\theta_0$ 
dependent provided that the vacuum mixing angle is not too close to $\pi/4$. 
This follows from the fact that $\sin^2 2\theta_0$ is the common factor in 
(\ref{P5}). 

(iv) If the phase $\phi_1=\delta\!\cdot\!t_1$ is not very close to $\pi/2$, 
the transition probability itself (i.e. neglecting matter effects) is also 
considerably enhanced as it now contains an $\sim \delta\!\cdot\!t$ term in 
addition to the $\sim (\delta\!\cdot\!t)^2$ ones. The relative size of
the matter effects in this case is  
\be
\cos 2\theta_0\tan\phi_1 (Vt)\,, 
\label{re2}
\ee
while for $\phi_1\approx \pi/2$ it is  
\be
\cos 2\theta_0 \,(V/\delta)\,. 
\label{re3}
\ee
Thus for generic values $s_1, c_1 \sim 1$ the contribution of the matter
effects to the transition probability, though strongly enhanced compared 
to the case of pure flavour eigenstate entering the matter, is relatively
small. It may, however, still be noticeable as the short pathlength  
approximation applies even for $Vt$ as large as $\sim 1/3$ (the
corrections are of the order $(Vt)^2$).  
When $\phi_1\simeq \pi/2$, matter effects give an important contribution 
to the transition probability over the time period $t$; they dominate in the 
limit $V \gg \delta$. This fact can be used for experimental searches of 
matter effects in short-pathlength neutrino experiments. 

(v) The estimates (\ref{re2}) and (\ref{re3}) apply to the situations when 
the probability of finding $\nu_b$ at the vacuum-matter border is either 
experimentally known (two-detector experiments), or can be reliably estimated 
theoretically. If this is not the case, one has to compare the matter-induced 
contributions to $P(\nu_i\to \nu_b)$ with the 
probability of finding $\nu_b$ in the final state 
itself rather than with the increase of this probability due to neutrino 
propagation in matter. The relative contribution of matter effects is then 
\be
\cos 2\theta_0 \,(V\!\delta\!\cdot\!t^2)\,.
\label{re4}
\ee
For $Vt\sim\delta\!\cdot\!t\sim 1/3$ it can be about 10\%. 

(vi) Eq. (\ref{P5}) is also valid when neutrinos first propagate a distance
$t$ in matter and then a distance $t_1$ in vacuum. 
This is a consequence of the fact that two-flavour neutrino oscillations in 
matter are invariant under time reversal {\em even if the matter density 
profile is not $T$ invariant.} Indeed, let the evolution matrix for eq. 
(\ref{Sch}) be $U(t_2, t_1)$, so that $|\nu(t_2)\rangle = U(t_2, t_1)|\nu(t_1)
\rangle$. The time-reversed evolution matrix is $U(t_1, t_2)=U(t_2, t_1)^{-1}= 
U(t_2, t_1)^\dag$. Since for any unitary $2\times 2$ matrix $|U_{21}| = 
|U_{12}|$, the probability of neutrino flavour oscillations is $T$ invariant. 
Notice that this is not in general true in the case of oscillations
between $n>2$ neutrino species.

Consider now a few numerical examples. 

Atmospheric neutrinos coming to a detector 
from below the horizon propagate first in the air (which for the purposes 
of neutrino oscillations can be considered as vacuum) and then in the
matter of the earth. The distances that neutrinos travel in the atmosphere
$t_1$ and in the earth $t$ are given by 
\be
t_1=-R |\cos\Theta|+\sqrt{(R+h)^2-R^2 \sin^2\Theta}\,,\quad\quad
t=-2R\cos\Theta\,,
\label{dist} 
\ee
where $R=6371$ km is the radius of the earth, $\Theta$ is the zenith angle of 
the neutrino trajectory, and $h\simeq 15$ km is the average hight at which the 
neutrinos are produced in the atmosphere. We are interested in the regime $t_1 
\gg t$ which corresponds to nearly horizontal neutrino trajectories (zenith 
angles only slightly exceeding $\pi/2$). In this case one can have sizeable 
phases $\phi_1$ whereas neutrino oscillations in the earth are in the
short pathlength regime. This corresponds to $|\cos\Theta|\aprle 0.01$.  
Atmospheric neutrinos with nearly horizontal trajectories pass through the 
earth's crust where the density is nearly constant and equal to about 2.8
$g/cm^3$ and the electron number fraction $Y_e\simeq 0.49$ (the same is
also true for short pathlength accelerator neutrino experiments). For
oscillations between active neutrinos $\nu_\mu\leftrightarrow\nu_e$ 
or $\nu_e\leftrightarrow\nu_\tau$ this gives $V\simeq 5.17\times 10^{-14}$
eV; for active-sterile neutrino oscillations $V$ is a factor of two smaller. 

Consider, e.g., atmospheric $\nu_\mu\leftrightarrow \nu_e$ (or $\nu_e 
\leftrightarrow \nu_\tau$) oscillations for the neutrino trajectory with the 
zenith angle $\Theta\simeq 1.581$ ($\cos\Theta=-0.01$). From eq. (\ref{dist}) 
one finds $t_1\simeq 378$ km, $t\simeq 127$ km, which gives $V t\simeq 
3.3\times 10^{-2}$. For $\Delta m^2=3.2\times 10^{-3}$ eV$^2$, which is
the current best-fit value of the Super-Kamiokande atmospheric neutrino
data \cite{SK1}, and $E=1$ GeV, 
one has $\delta=8\times 10^{-13}$ eV, $\phi_1\simeq 1.535$, $\delta\cdot 
t \simeq 0.517$. The relative matter contribution to the total transition 
probability is 4.3\% and that to the increase of the probability due to  
neutrino propagation in the earth is 18\%. This is a large effect, taking 
into account that the distance neutrinos travel inside the earth is only 
127 km. If neutrinos did not oscillate in the atmosphere before entering the 
earth, this contribution would have been only about 0.5\%. 
It should be noted, however, that the short-pathlength regime is valid only
in a narrow range of zenith angles. In addition, the effective mixing angle 
for $\nu_e\leftrightarrow \nu_x$ oscillations is known to be small
\cite{chooz}. 

Oscillations of $\nu_\mu$ into sterile neutrinos are disfavoured as the 
dominant channel of the atmospheric neutrino oscillations \cite{SK2}, but 
allowed as a subdominant channel with a weight that can be as large as about 
50\% \cite{Lisi}. For this channel, relative matter effects are a factor
of two smaller than they are for the $\nu_\mu\leftrightarrow \nu_e$ or 
$\nu_e \leftrightarrow \nu_\tau$ channels, but their absolute value can be 
significantly larger because the corresponding mixing angle can be quite
large. 

Consider now the situation when neutrinos first propagate in matter and 
then in vacuum. This could, e.g., be realized in accelerator neutrino 
experiments with a detector placed on an earth's satellite, which is 
certainly a rather remote possibility. Let us assume for definiteness that 
the distance that neutrinos propagate in the earth $t=730$ km (the baseline 
of CERN -- Gran Sasso and Fermilab -- Soudan mine experiments). This 
corresponds to $\cos\Theta=-5.7344\times 10^{-2}$, $V t\simeq 0.191$ (for 
oscillations between active neutrinos). Assuming that the height of the 
satellite's orbit is 750 km, the distance that neutrinos propagate after
exiting the earth $t_1\simeq 2837$ km. Let us again take $\Delta m^2=3.2\times 
10^{-3}$ eV$^2$. For $E=8$ GeV, which is a typical energy of the accelerator 
neutrino experiments, one has $\delta=10^{-13}$ eV, $\phi_1\simeq 1.438$, 
$\delta\cdot t \simeq 0.37$. The relative matter contribution to the total 
transition probability is 13\%, and that to the transition probability
acquired due to neutrino propagation in the earth is a factor 
1.3. This means that matter effects dominate the transition inside the
earth in this case. If the detector is placed on the surface of the earth,
i.e. $t_1=0$, matter effects constitute only about 3.5\% of the oscillation 
probability for neutrinos that have traversed the earth, i.e. are almost a
factor of 37 smaller. 
Thus, as paradoxical as it looks, the earth's matter effects on the
oscillation probability become stronger when neutrinos are detected
farther from the surface of the earth. 

Can one achieve a significant enhancement of the matter effects in the 
accelerator experiments by having very long decay tunnels? Unfortunately, 
this does not seem to be possible since a sizeable enhancement is only
achieved when the distance traveled by neutrinos in vacuum is much larger 
than their pathlength in matter, and the latter should be at least a few
hundred km.

\section{Day-night effect in solar neutrino experiments}

Another example of an initial state which is not a flavour eigenstate 
is given by the earth matter effects on solar neutrinos coming to a detector 
during night in the case of the MSW \cite{W,MS} solutions of the solar 
neutrino problem. In this case the neutrino 
state arriving at the earth is an incoherent superposition of the mass 
eigenstates $\nu_1$ and $\nu_2$ (see, e.g., \cite{DLS} for a recent 
discussion). The probability of finding a $\nu_e$ at the detector depends on 
the probability of $\nu_2\to\nu_e$ oscillations inside the earth $P_{2e}$. 
Since in this case the initial state is a mass eigenstate $\nu_2$, one has  
$a_1=\sin \theta_0$, $a_2=\cos \theta_0$. In the short pathlength limit, in 
the case of matter of constant density, one finds from (\ref{P3a})  
\be
P_{2e}-(P_{2e})_{init} = P_{2e}-\sin^2\theta_0 = \sin^2 2\theta_0\, 
(V\!\delta\!\cdot\!t^2)\,.
\label{P2e}
\ee
This expression vanishes in the $V\to 0$ limit because mass eigenstates do 
not oscillate in vacuum. However, it is nontrivial that it is of the order 
$V\!\delta\!\cdot\!t^2$ rather than $(V\!\delta\!\cdot\!t^2)^2$; because of 
this the day-night effect in solar neutrino experiments can be sizeable 
and must not be neglected even when the neutrino pathlengths inside the
earth are relatively short. Since $a_1$ and $a_2$ are both real, there is no 
$\sim \delta\!\cdot\!t$ contribution  to the transition probability. 

Another interesting point to notice is that the r.h.s. of (\ref{P2e}) 
is $\propto 1/E$. For sizeable pathlengths, the day-night effect increases 
with neutrino energy when $\delta \gg V$ and decreases when $\delta \ll V$ 
(see, e.g., fig. 4 in \cite{BK}). From (\ref{P2e}) it follows that for short 
pathlengths it always decreases with $E$, irrespective of the relative 
magnitudes of $V$ and $\delta$. 

Let us consider a few numerical examples. For $Vt\sim \delta\!\cdot\!t\sim
1/3$, which corresponds to the neutrino pathlength inside the earth
$t\simeq 1270$ km and $\delta\simeq V\simeq 5.2\times 10^{-14}$ eV,  
and assuming $\sin^2 2\theta_0\simeq 1$, the matter-induced oscillation 
probability (\ref{P2e}) is of the order of 10\%. For $^7$Be solar neutrinos 
($E=0.862$ MeV) the above value of $\delta$ corresponds to $\Delta m^2 
\simeq 1.8 \times 10^{-7}$ eV$^2$, 
which is in the range of the MSW-LOW solution of the solar neutrino problem 
(for a recent analysis of the solar neutrino data see \cite{Valencia}). For 
the same distance $t\simeq 1270$ km, matter effects on the probability of 
neutrino flavor oscillations would constitute only about 1\%, i.e. an 
order of magnitude smaller. For typical parameters of the MSW-LMA solution 
and $E\simeq 10$ MeV (typical energy of the $^8$B solar neutrinos) 
eq.~(\ref{P2e}) yields the probability $\sim 1$\% for the pathlengths as
short as about 100 km, which has to be compared with the value $10^{-4}$ 
for the earth matter effect on neutrino flavour oscillations over the 
distance of 100 km. 

Similar considerations apply to oscillations of supernova neutrinos 
inside the earth since those neutrinos also arrive at the earth as mass 
eigenstates. If the next supernova explosion occurs at such a time that
its neutrinos come to a terrestrial detector passing through a short 
distance inside the earth, the earth's matter effects on their oscillations 
may still be quite strong and should be taken into account
\footnote{The importance of matter effects on oscillations of supernova 
neutrinos inside the earth was recently emphasized in \cite{SN}; 
however, the short pathlength limit was not discussed in these papers.}.

Mass eigenstate neutrinos are also produced in neutral current 
reactions, i.e. in decays of real or virtual $Z^0$ bosons; it appears, 
however, technically rather difficult to produce sizeable beams of 
neutrinos born in neutral current reactions. 

\section{Neutrino oscillations due to FCNC}

Neutrinos can oscillate in matter even if their masses are zero or
negligible, provided that they have flavour changing neutral current (FCNC) 
interactions \cite{W}. The evolution of the system is described by 
eq. (\ref{Sch}) with $A$ and $B$ being now both $t$ dependent:
\be
A(t)=\frac{G_F}{\sqrt{2}}\,\epsilon'\,N(t)\,, \quad\quad 
B(t)=\sqrt{2}{G_F}\,\epsilon\,N(t)\,, 
\label{FCNC}
\ee
Here $\epsilon$ and $\epsilon'$ are the FCNC parameters. 
In the short-baseline approximation, the transition probability 
$P(\nu_a\to \nu_b;\,t)\simeq B^2 t^2$. In contrast to the case of the 
ordinary neutrino oscillations in which the off-diagonal element $B$ of the 
effective Hamiltonian is independent of matter density $N(t)$, in the case 
of oscillations due to FCNC it is proportional to $N(t)$. Therefore the 
matter effect on the oscillation probability (which in this case coincides  
with the probability itself) is of the order $(V t)^2$ rather than $(V\!\delta
\!\cdot\!t^2)^2$. This explains why in this case, unlike in the case of
the ordinary neutrino flavour oscillations, matter effects 
can be quite sizeable even for as short baselines as that of the K2K 
experiment, $t=250$ km (this fact was previously pointed out in \cite{OP,LS}). 
The expression $P(\nu_a\to \nu_b;\,t)\simeq B^2 t^2$ also explains why in the 
short baseline limit matter effects do not depend on $\epsilon'$ in the 
leading order \cite{OP}.

\section{Summary and conclusion}

We discussed matter effects in short-pathlength neutrino oscillations and
found out the reason why these effects generally decrease with decreasing
pathlength more rapidly than the oscillation effects themselves. This happens 
because, in the leading order in perturbation theory, the transition 
amplitudes are determined by the off-diagonal terms of the effective
Hamiltonian which are matter independent. This, however, is not true if 
the initial neutrino state is not a flavour eigenstate, in which case 
the diagonal terms of the effective Hamiltonian also contribute, or if the 
oscillations are due to flavour changing neutral currents, when the 
off-diagonal terms depend on matter density. 

Initial states which are not flavour eigenstates can be obtained if neutrinos 
propagate significant distances in vacuum before entering matter (or after
exiting it). In this case matter effects in short-pathlength neutrino 
oscillations can be strongly enhanced. 
We discussed implications of this observation for atmospheric neutrinos
with nearly horizontal trajectories and for short pathlength accelerator 
experiments with the detector placed at large distances from the point 
where neutrinos exit the earth. 

One also deals with initial neutrino states which are not flavour
eigenstates when considering the earth matter effects on solar neutrinos in 
the case of the MSW solutions of the solar neutrino problem, or oscillations 
of the supernova neutrinos inside the earth. In these cases the initial states 
are mass eigenstate neutrinos. Matter effects for short neutrino pathlengths 
inside the earth are of the order $V\!\delta\!\cdot\!t^2$ rather than
$(V\!\delta\!\cdot\!t^2)^2$ which would be expected if the initial state  
were flavour eigenstate. 

In conclusion, we have shown that matter effects in oscillations of neutrinos 
traveling short distances in matter can be strongly enhanced when the initial 
neutrino state entering the matter is not a flavour eigenstate, or when 
neutrinos are detected at significant distances from the point where they exit 
the matter. We believe that this is an interesting observation, even though 
the conditions for such an enhancement are generally difficult to realize. 
A notable exception is provided by solar and supernova neutrinos for which 
these conditions are realized as they arrive at the earth in mass rather
than in flavour eigenstates. 

The author is grateful to H. Minakata for asking a question during the 
NOW2000 workshop which led to the present study, and to A. Yu. Smirnov 
for useful discussions. 
This work was supported by Funda\c{c}\~ao para a Ci\^encia e a Tecnologia
through the grant JNICT-CERN/P/FIS/15184/99 and by the TMR network grant
ERBFMRX-CT960090 of the European Union.

\end{document}